\documentclass{article}
\usepackage{LQGNet}

\title{LQGNet: Hybrid Model-Based and Data-Driven Linear Quadratic Stochastic Control}
\name{Solomon Goldgraber Casspi, Oliver H\"{u}sser, Guy Revach, and Nir Shlezinger
\thanks{S. Goldgraber Casspi and N. Shlezinger are with the School of ECE, Ben-Gurion University of the Negev, Be`er Sheva, Israel (e-mail: casspi@post.bgu.ac.il; nirshl@bgu.ac.il).
 O. H\"{u}sser and G. Revach are with the D-ITET, ETH Zurich, Switzerland, (email: huesseol@student.ethz.ch; grevach@ethz.ch). We thank Hans-Andrea Loeliger for helpful discussions.}}

\address{\vspace{-25mm}}
\begin{document}
%\ninept
%
\maketitle
%
%%%%%%%%%%%%%%%%
%%% Abstract %%%
%%%%%%%%%%%%%%%%
%
% Stochastic control plays a key role in numerous applications. A common framework for designing control policies utilizes 
% quadratic objectives while representing the dynamics as a known linear Gaussian \ac{ss} model. 
% For this \ac{lqg} setting, the optimal controller is obtained in closed-form by the separation principle. However, in practice, the underlying dynamics are often not faithfully captured by a linear Gaussian \ac{ss} model, limiting the performance. Here, we present LQGNet, a stochastic controller that leverages data to operate under partially known dynamics. LQGNet augments the state tracking module of separation-based control with a dedicated trainable algorithm. The resulting system preserves the operation of classic \ac{lqg} control while learning to cope with partially known \ac{ss} models without having to fully identify the dynamics.  We numerically show that LQGNet outperforms classic stochastic control by overcoming mismatched \ac{ss} models.
%
%%%%%%%%%%%%%%%%
%%% Abstract %%%
%%%%%%%%%%%%%%%%
%
\begin{abstract}

Stochastic control deals with finding an optimal control signal for a dynamical system in a setting with uncertainty,  playing a key role in numerous applications. The \ac{lqg} is a widely-used setting, where the system dynamics is represented as a \acl{lg} \ac{ss} model, and the objective function is quadratic. For this setting, the optimal controller is obtained in closed form by the separation principle. However, in practice, the underlying system dynamics often cannot be faithfully captured by a {fully known} \acl{lg} \ac{ss} model, limiting its performance. Here, we present \acl{ln}, a stochastic controller that leverages data to operate under partially known dynamics. \acl{ln} augments the state tracking module of separation-based control with a dedicated trainable algorithm. The resulting system preserves the operation of classic \ac{lqg} control while learning to cope with partially known \ac{ss} models without having to fully identify the dynamics. We empirically show that \acl{ln} outperforms classic stochastic control by overcoming mismatched \ac{ss} models.
\end{abstract}
\acresetall
%
%%%%%%%%%%%%%%%%%%%%
%%% Introduction %%%
%%%%%%%%%%%%%%%%%%%%
%
%\vspace{-0.2cm}
\section{Introduction}\label{sec:intro}
{Stochastic optimal control is a sub-field of mathematical optimization with applications spanning from operations research to physical sciences and engineering, including aerospace, vehicular systems, and robotics~\cite{OCTH_1950_1985}.}
%
% Optimal control theory is a mathematical discipline with numerous applications in science and engineering. The origin of optimal control theory date back to the 17th century with the calculus of variations and continued to develop during the 18th-20th century, having captured the interest and attention of the greatest minds during those times. Optimal control theory truly started to be realized in engineering during the 1950s, when the digital computer became available commercially \cite{OCTH_1950_1985}. 
% \par 
% Today, control theory is integrated into many modern applications, such as aircraft control, automobiles, satellites, robotics, and industrial processes. The dynamics of physical processes and sensor measurements are natively random. Therefore, in the context of control theory, stochastic control is a fundamental approach to real-world tasks. 
% %
% \subsection{Relevant literature}
%
% Stochastic control focuses on dynamic systems with noisy sensory data. A fundamental stochastic setting is  \ac{lqg} control \cite{Athans}, where the dynamics obey a linear Gaussian \ac{ss} model, and the controller should minimize a quadratic objective. For such settings, the optimal  controller follows the separation principle \cite{YOSHIKAWA1978, gunckel1963general}, where state estimation is decoupled from control. 
%
Stochastic control consdiers dynamical system under the existence of uncertainty, either in its evolution or in its observations.The aim is to find an optimal control signal for a given objective function.
%
% A fundamental stochastic setting is  \ac{lqg} control \cite{Athans}, where the dynamics obey a linear Gaussian \ac{ss} model, and the controller should minimize a quadratic objective. For such settings, the optimal  controller follows the separation principle \cite{YOSHIKAWA1978, gunckel1963general}, where state estimation is decoupled from control.
%
%
In the fundamental \ac{lqg} setting \cite{Athans}, e the system dynamics obey a \acl{lg} \ac{ss} model, and the controller should minimize a quadratic objective. The optimal \ac{lqg} controller follows the separation principle \cite{YOSHIKAWA1978, gunckel1963general}, where state estimation is decoupled from control, and it comprises a \ac{kf} followed by a conventional \ac{lqr}~\cite{Gunckel1963}.

While {\ac{lqg} control} is simple and tractable, it relies on the ability to faithfully describe the dynamics as a  fully known \acl{lg} \ac{ss} model. In practice, \ac{ss} models are often approximations of the system's true dynamics, while its stochasticity can be non-Gaussian. The presence of such mismatched domain knowledge notably affects the performance of classical \acl{mb} policies.

To overcome the drawbacks of oversimplified modeling, one can resort to learning. The main learning-based approach in sequential decision making and control is \ac{rl}~\cite{BertsekasT96, sutton2018reinforcement} where an agent is trained via experience-driven autonomous learning to maximize a reward \cite{SILVER2021}. The growing popularity of model-agnostic \acp{dnn} %for problem solving without the need to explicitly characterize the processes governing the domain of interest ~\cite{bengio2009learning},
and their  empirical success in various tasks involving complex data, such as visual and language data, has led  to a growing interest in
deep \ac{rl}~\cite{ArulkumaranDBB17}. %in various tasks, ranging from games~\cite{silver2018general} to algorithm discovery~\cite{fawzi2022discovering}.} 
Deep \ac{rl} systems based on black-box \acp{dnn} were proposed for implementing controllers for various tasks, including robotics and vehicular systems~\cite{kuutti2020survey,zhang2022survey}. Despite their success, these architectures cannot naturally incorporate the domain knowledge available in partially known \ac{ss} models, are  complex and difficult to train, and lack the interpretability of \acl{mb} methods~\cite{shlezinger2020model}.
%
%to using deep learning for control
{An alternative approach} uses \acp{dnn} to extract features processed with model-based methods \cite{iwata2021controlling,peralez2021deep,perez2021deep}. This approach still requires one to impose a fully known \ac{ss} model of the features, motivating the incorporation of deep learning  into classic controllers to bypass the need to fully characterize the dynamics. 
{In this work, we propose \acl{ln}, a hybrid  stochastic controller designed via model-based deep learning \cite{shlezinger2020model,shlezinger2022model}. \acl{ln}  preserves the structure of the optimal \ac{lqg} policy, while operating in partially known settings.
We adopt the recent \acl{kn} architecture~\cite{KalmanNet}, which implements a trainable \ac{kf}, in a separation-based controller. The resulting \acl{ln} architecture utilizes the available domain knowledge by exploiting the system's description as a \ac{ss} model, 
thus preserving the simplicity and interpretability of the \acl{mb} policy while leveraging data to overcome partial information and model mismatches. By converting the optimal \acl{mb} \ac{lqg} controller into a trainable discriminative algorithm \cite{shlezinger2022discriminative} our \acl{ln} \emph{learns to control} in an \acl{e2e} manner. We empirically demonstrate that \acl{ln} approaches the performance of the optimal \ac{lqg} policy with fully known \ac{ss} models, and notably outperforms it in the presence of mismatches.} 

The rest of this paper is organized as follows: 
\secref{sec:format} reviews the \ac{ss} model and {details} the \ac{lqg} control {task.} \secref{sec:lqgnet} presents \acl{ln}, which is  evaluated in \secref{sec:Results}. %\secref{sec:Conclusions} provides concluding remarks.
% \par
% \textcolor{red}{Nir, in this part you suggested including a detailed description of what we do. Details are already provided in section 3, and specifically in subsection 3.2. "Training Algorithm". Should we provide details anyway? or maybe move the already detailed explanation from next section to here?}

% \subsection{Notations and Preliminaries}
% We use $x$, $\mathbf{x}$ and $\mathbf{X}$ for a scalar, column vector and matrix,
% respectively. The transpose, inverse, $\ell_2$ norm, and stochastic expectation are denoted by $\set{\cdot}^\top$, $\set{\cdot}^{-1}$, $\norm{\cdot}$, and  $\expecteds{\cdot}$, respectively. Finally, $\greal$ is the set of real numbers.
%
%%%%%%%%%%%%%%%%%%%%%%%%%%%%%%%%%%%%%%%%%%%%%%%%
%%% System Model & Preliminaries (Section 2) %%%
%%%%%%%%%%%%%%%%%%%%%%%%%%%%%%%%%%%%%%%%%%%%%%%%
%\vspace{-0.2cm}
\section{System Model and Preliminaries}\label{sec:format}
%\vspace{-0.1cm}
%
As a preliminary step to deriving \ac{ln}, we describe in the 
 describes the system model in \ssecref{ssec:model}.  \ssecref{ssec:task} the formulates the \acl{dd} \ac{lqg} control task in partially known \ac{ss} models, and \ssecref{ssec:lqg}  reviews basics in optimal \acl{mb} \ac{lqg} control.
%
% \subsection{State-Space Model}
% The state space (SS) representation of a system replaces an $n$th order differential equation with a single first order matrix differential equation. The general SS representation of a discrete-time system is given by two equations:
% %
% \begin{subequations}
%     \begin{align}
%         \x_{t} & = \mathbf{f}(\x_{t-1}, \u_{t-1}, \w_t), &\x_t &\in \mathbb{R}^{n}\label{eq:discrete_ss_model_a},\\
%         \y_t & =  \mathbf{h}(\x_t, \v_t), &\y_t &\in \mathbb{R}^{m}\label{eq:discrete_ss_model_b}.
%     \end{align}   
% \end{subequations}
% %
% At discrete time steps $t\in \mathbb{W}$, $\x_{t}$ is the evolved state from the previous state $\x_{t-1}$, $\u_t \in \mathbb{R}^{q}$ is the input control, $\w_t \in \mathcal{W}$ is the system disturbance, and $\mathbf{f}(\cdot)$ is possibly nonlinear state-evolution function. $\y_t$ is the vector of observations that is generated from the current state vector at time step $t$ by possibly nonlinear observation mapping $\mathbf{h}(\cdot)$ corrupted by measurement noise $\v_t \in \mathcal{V}$.
%
%%%%%%%%%%%%%%%%%%%%
%%% System Model %%%
%%%%%%%%%%%%%%%%%%%%
%
%\vspace{-0.1cm}
\subsection{Dynamical System Model}\label{ssec:model}
%\vspace{-0.1cm}
% Such models describe the relationship at each time instance $t$ between a state vector  $\x_t \in \mathbb{R}^{m}$, an input control signal $\u_t \in \mathbb{R}^{q}$, and the observations $\y_t \in \mathbb{R}^{n}$. We focus on \ac{ss} models with a linear  state-evolution 
% %
% \begin{subequations}
% \label{eq:assumed_problem}
%     \begin{align}
%         \x_{t} & = \F \x_{t-1}+\G \u_{t-1} + \w_t, %&\w_t \sim \mathcal{N}(\mathbf{0},\mathbf{W})
%         \label{eq:assumed_problem_a}
%     \end{align}    
% for some fixed  matrices $\F \in \mathbb{R}^{m \times m},\G \in \mathbb{R}^{m \times q}$ and zero-mean noise $\w_t \in \mathbb{R}^m$. The observations are given by
%     \begin{align} 
%         \y_t & = \H\x_t + \v_t,\label{eq:assumed_problem_b}
%     \end{align}    
% \end{subequations}
% %
% where $\H \in \mathbb{R}^{n \times m}$ is the observation (emission) mapping or sensing function, and $\v_t\in  \mathbb{R}^{n}$ is a zero-mean noise signal.
%
{We consider dynamic systems characterized by a \ac{ss} model in discrete-time $t\in\gint$. This \ac{ss} model describes the relationship between a state vector $\x_t \in \greal^{m}$, an input control signal $\u_t \in \greal^{q}$, and the noisy observations $\y_t\in\greal^{n}$, at each time instance $t$. We focus on \acl{lg} models, given by
\begin{subequations}\label{eq:SS_model}
\begin{align}\label{eqn:stateEvolution}
\gvec{x}_{t}&= 
\F\cdot\x_{t-1}+\G\cdot\u_{t-1}+\w_t,
&\gvec{w}_t\sim
\gnormal{\gvec{0},\gvec{W}},\\ \label{eqn:stateObservation}
\gvec{y}_{t}&=
\H\cdot\gvec{x}_{t}+\gvec{v}_{t},
&\gvec{v}_t\sim
\mathcal{N}\brackets{\gvec{0},\gvec{V}}.
\end{align}
\end{subequations}
 Here, $\F\in\greal^{m\times m}$, $\G\in\greal^{m\times q}$, and $\H\in \greal^{n\times m}$ are the evolution, control, and observation (emission) matrices, respectively, and $\w_t,\v_t$ are \acl{awgn} signals with covarainces $\gvec{W}, \gvec{V}$, respectively.}
%
%%%%%%%%%%%%%%%%%%%%%%%%%%%%%%%%%%%%%%%%%%%
%%% Data Driven Stochastic Control Task %%%
%%%%%%%%%%%%%%%%%%%%%%%%%%%%%%%%%%%%%%%%%%%
%
\subsection{Data Driven Stochastic Control Task}\label{ssec:task}
The stochastic control {task} considers the {optimization} of a policy $\psi:\greal^{n}\mapsto\greal^{q}$, which maps {a noisy observation signal} $\y_t$ into a control signal $\u_t$, under 
quadratic control loss. Over a finite-horizon of $\gscal{T}$ time steps, the loss is given by
\begin{align}\label{eq:LQR_cost}
\J(\psi) = 
\tilde{\x}_\gscal{T}^{\top}\gvec{Q}_\gscal{T}\tilde{\x}_\gscal{T}
+\sum_{t=0}^{\gscal{T}-1}
\brackets{\tilde{\x}_{t}^{\top}\gvec{Q}_{\tilde{\x}}\tilde{\x}_{t} 
+
\tilde{\u}_{t}^{\top}\gvec{R}_{\tilde{\u}}\tilde{\u}_{t}}.
\end{align}
The loss \eqref{eq:LQR_cost}
balances the stability of the closed-loop system and the aggressiveness of control. Here,  $\tilde{\u}_{t}$ and $\tilde{\x}_{t}$ are the deviations of $\u_t$ and $\x_t$, respectively,  from predefined target values. The former is typically set to zero, while the latter is a desired state that depends on the regulation problem. The matrices $\gvec{Q}_{\tilde{\x}},\gvec{Q}_{\gscal{T}}\succeq0$, and $\gvec{R}_{\tilde{\u}}\succ0$ are predefined weighting costs for state, final state, and input control, respectively. While we assume to have access to a (possibly mismatched) estimate of the \ac{ss} \emph{design} matrices $\F,\G$, $\H$, we do not assume prior knowledge of the distribution of the noises. 

To overcome this missing domain knowledge, we are given access to a simulator $\Gamma:\greal^q\mapsto\greal^{m}\times \greal^n$ that emulates the underlying dynamics. The simulator allows generating random trajectories of inputs while measuring their observations and state values, i.e., 
%
%\begin{equation}
$(\y_t,\x_t) = \Gamma(\u_{t-1})$.   
%\end{equation}
%
%%%%%%%%%%%%%%%%%%%
%%% Optimal LQG %%%
%%%%%%%%%%%%%%%%%%%
%
\subsection{Optimal LQG Control}\label{ssec:lqg}
%
% \ac{lqg} is one of the most fundamental optimal control frameworks \cite{Athans}. It considers linear control in \ac{ss} models as in \eqref{eq:assumed_problem} where: $1)$ the noise signals $\w_t$ and $\v_t$ are both i.i.d. and Gaussian with covarinces $\W$ and $\V$, respectively;  $2)$ the parameters of the \ac{ss} models are fully known. Under such settings, the separation principle of stochastic control \cite{YOSHIKAWA1978,Gunckel1963} reveals that the control policy $\psi^{\rm LQG}$ which minimizes \eqref{eq:LQR_cost} can be decomposed into two separate modules: {\em \ac{mse}-optimal state estimate} followed by {\em full-state optimal \ac{lqr}}.
%
\ac{lqg} is one of the most fundamental optimal control frameworks \cite{Athans}. It considers dynamics with a \acl{lg} \ac{ss} model as in \eqref{eq:SS_model} with a quadratic objective as in \eqref{eq:LQR_cost}. For such setting, \eqref{eq:SS_model}, the \emph{separation principle}~\cite{YOSHIKAWA1978,Gunckel1963} applies, and the optimal policy $\psi^\ast$ which minimizes \eqref{eq:LQR_cost} can be decomposed into two separate modules: a \ac{mse} optimal state estimator, namely the \ac{kf}, followed by full-state optimal \ac{lqr}. Both modules, detailed next, require full knowledge of the \ac{ss} model parameters.  
%
% {\bf \ac{mse} optimal state estimator}: 
% The first part of the optimal \ac{lqg} controller is the minimal \ac{mse} estimate of $\x_t$, denoted $\hat{\x}_t$, from $\{\y_\tau\}_{\tau\leq t}$. This estimator is realized by the \ac{kf}, which at each time  $t$ first predicts the state and observations based on the previous estimate $\hat{\x}_{t-1}$ via $\hat{\x}_{t|t-1}=\F \hat{\x}_{t-1} + \G \u_{t-1}$ and $\hat{\y}_{t|t-1}=\H \hat{\x}_{t|t-1}$.
% Then, it tracks the covariances of these predictions via 
% ${\boldsymbol{\Sigma}}_{t|t-1} = \F {\boldsymbol{\Sigma}}_{t-1} \F^\top + \W$ and $\S_t = \H {\boldsymbol{\Sigma}}_{t|t-1} \H^\top + \V$.
% The next estimate is then updated as
% \begin{equation}
%     % \hat{\x}_{t}=\K_t (\y_t - \hat{\y}_{t|t-1}) + \hat{\x}_{t|t-1},
%     \hat{\x}_{t}=\K_t \Delta\y_t + \hat{\x}_{t|t-1},
% \end{equation}
% where $\Delta\y_t = \y_t - \hat{\y}_{t|t-1}$ is the innovation, and $\K_t = {\boldsymbol{\Sigma}}_{t|t-1} \H^\top {\S}_{t}^{-1}$ is the Kalman gain. The covariance of this estimate is updated as ${\boldsymbol{\Sigma}}_{t} = {\boldsymbol{\Sigma}}_{t|t-1} - \K_t \S_t \K_t^\top$.
% 
%\vspace{-0.2cm}
\subsubsection{Kalman filter - \ac{mse} optimal state estimator}
{The \ac{kf} is an efficient linear recursive state estimator, which for every time step $t$ produces an estimate $\hat{\x}_t$ for $\x_t$, based on all previous and current observations $\{\y_\tau\}_{\tau\leq t}$.
It can be described as a two-step procedure; prediction and update, that uses only the new observation and the previous estimate as sufficient statistics to compute first and second order moments. The prediction is given by
\begin{subequations}\label{eq:predict}
\begin{align}
\label{eq:predictx}
\hat{\gvec{x}}_{t\given{t-1}}&\!=\! 
\gvec{F}{\hat{\gvec{x}}_{t-1}}\!+\!\G\u_{t-1},
\hspace{0.2cm}
\mySigma_{t\given{t-1}}\!=\!
\gvec{F}\mySigma_{t-1}\gvec{F}^\top\!+\!\gvec{W},\\
\label{eq:predicty}
\hat{\gvec{y}}_{t\given{t-1}}&\!=\!
\gvec{H}{\hat{\gvec{x}}_{t\given{t-1}}},
\hspace{1.35cm}
\gvec{S}_{t}\!=\!
\gvec{H}\mySigma_{t\given{t-1}}\gvec{H}^\top\!+\!\gvec{V},
\end{align}
\end{subequations}
and the update is given by
%
%%%%%%%%%%%%%%%%%%%%%%%
%%% Update Equation %%%
%%%%%%%%%%%%%%%%%%%%%%%
%
\begin{subequations}
\begin{align}\label{eq:update}
\hat{\gvec{x}}_{t}&= 
\hat{\gvec{x}}_{t\given{t-1}}+\K_t\cdot\Delta\gvec{y}_t,
\quad
\Delta\gvec{y}_t=\gvec{y}_t-\hat{\gvec{y}}_{t\given{t-1}},\\
\K_{t}&={\mySigma}_{t\given{t-1}}{\gvec{H}^\top}{\gvec{S}}^{-1}_{t}, 
\quad
{\mySigma}_{t}\!=\!
{\mySigma}_{t\given{t-1}}\!-\!\K_{t}{\mathbf{S}}_{t}
\K^{\top}_{t}. \label{eq:updateKG}
\end{align}
\end{subequations}
Here, $\K_t$ is the \ac{kg}, computed recursively based on tracking the second-order moments of the signals. %, and $\Delta\y_t$ is the innovation term.
}
%
%\vspace{-0.2cm}
\subsubsection{Optimal LQR policy}
%{\bf Optimal \ac{lqr} policy}: 
For the system in \eqref{eq:SS_model} with known $(\F,\G)$, the optimal control input is given by 
\begin{equation} \label{eq:Optimal_control}
\u_t \!=\! -\L_t\hat{\x}_t,
\hspace{0.3cm}
\L_t \!=\! \left[\R_{\tilde{\u}}+\G^\top \P_{t+1} \G\right]^{-1} \G^\top \P_{t+1} \F.
\end{equation}
$\L_t$ is control gain and $\mathbf{P}_t$ is the solution to
the \acl{dare}~\cite{OptimalControl3rdEdition}, computed as 
%
% \begin{align}\label{eq:DARE}
% \P_{t-1} = &\;\Q_{\tilde{\x}} + \F^\top \P_t \F\notag\\ &-\F^\top \P_t \G \left[\R_{\tilde{\u}}+\G^\top \P_t \G\right]^{-1}\G^\top\P_t\F.
% \end{align} 
\begin{equation}\label{eq:DARE}
\P_{t-1} = \Q_{\tilde{\x}} + \F^\top \P_t \F-\F^\top \P_t \G \cdot\L_{t-1}.
\end{equation}
The above controller coincides with the \ac{lqr} policy, that minimizes the quadratic objective for \acl{lg} \ac{ss} models where one has a noise-free observation of the state. 
\vspace{-0.15cm}
\section{LQGNet}\label{sec:lqgnet}
%
%\vspace{-0.1cm}
Next, we present \acl{ln}, which learns to implement \ac{lqg} control under the considered partially known \ac{ss} model. We begin by detailing the architecture of \acl{ln} in \ssecref{HighLevelArchitecture}, after which we describe the training procedure and provide a discussion in Subsections~\ref{ssec:training}-\ref{ssec:discussion}, respectively. 

% Here, we present LQGNet, which is a hybrid model-based/data-driven stochastic control policy that is derived by augmenting the optimal \ac{lqg} policy with deep learning components.  We use KalmanNet \cite{KalmanNet}, a hybrid, interpretable, data-efficient architecture for real-time state estimation in nonlinear dynamical systems with partial domain knowledge, in the LQG problem. KalmanNet combines MB Kalman filtering with an RNN to cope with model mismatch
% and non-linearities. To introduce LQGNet, we explain its high-level operation in Subsection \ref{HighLevelArchitecture}.
% Then we present the features processed by its internal RNN and the specific architectures considered for implementing and training KalmanNet in Subsections III-BIII-
% D. Finally, we provide a discussion in Subsection III-E.
%\vspace{-0.1cm}
%
%%%%%%%%%%%%%%%%%%%%%%%%%%%
%%% LQGNet Architecture %%%
%%%%%%%%%%%%%%%%%%%%%%%%%%%
%
% state estimate produced by \acl{kn} is then processed by the \acl{mb} \ac{lqr} \eqref{eq:Optimal_control}, and thus the available (though possibly approximated) $\G$, $\F$, and $\H$ are utilized by the \ac{kf} predict step and for computing the control gain $\L_t$. The resulting architecture is illustrated in Fig. \ref{fig:LQGNet_scheme}.
%
\subsection{LQGNet Architecture}\label{HighLevelArchitecture}
The optimal \ac{lqg} controller detailed in Subsection~\ref{ssec:lqg} requiresfull access to the \ac{ss} model  \eqref{eq:SS_model}. While the \emph{design} matrices $\F,\G$, $\H$ are assumed to be partially known, the distributions of the noise signals are not known at all. Consequently, to cope with this missing knowledge without imposing a model and estimating the statistics of the noise {signals}, \acl{ln} augments the state estimator with deep learning. 

\begin{figure}
\centering
\includegraphics[width=0.95\columnwidth]{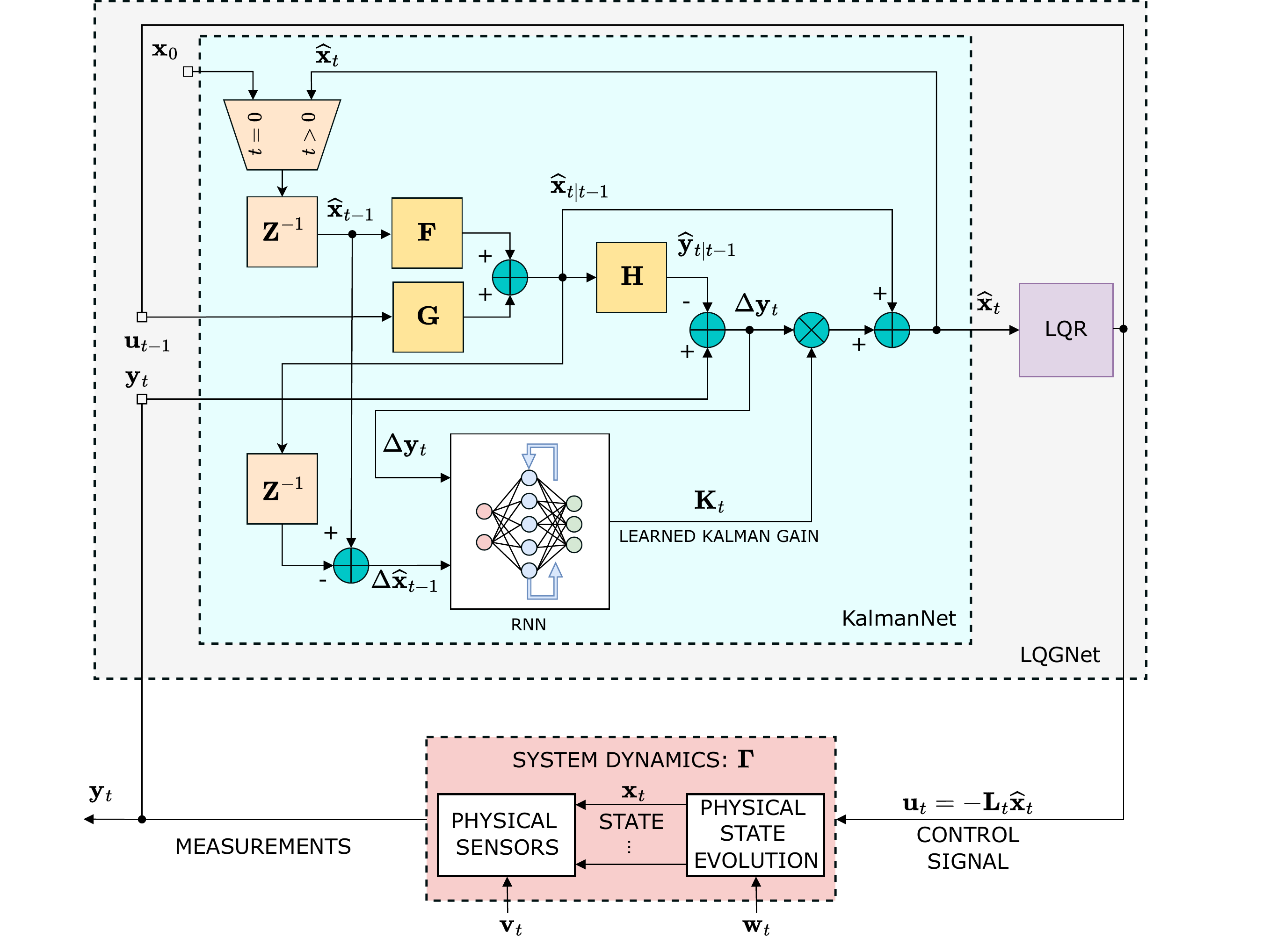}
%\vspace{-0.2cm}
\caption{LQGNet block diagram.}
\vspace{-0.15cm}
\label{fig:LQGNet_scheme}
\end{figure}

Since the \ac{lqg} optimal controller employs the \ac{kf} for state estimation, we replace it with our recently proposed \acl{kn} architecture~\cite{KalmanNet}. \acl{kn} is particularly suitable here, as it preserves the flow and interpretable nature of the \ac{kf}~\cite{klein2021uncertainty}. {While the \acl{mb} \ac{kf} requires the noise statistics to formulate the \ac{kg} $\K_t$ \eqref{eq:updateKG}, \acl{kn} uses a trainable \ac{rnn} to compute it, bypassing the need to impose a model on the noises. More specifically, \acl{kn} predicts the first order moments 
using the design matrices, as in \eqref{eq:predict}, and then updates using its learned (surrogate) \ac{kg}, as in \eqref{eq:update}.
The state estimate produced by \acl{kn} is then processed by the \acl{mb} \ac{lqr} \eqref{eq:Optimal_control} and the gain $\L_t$ using the available (though possibly approximated) design matrices, where the noise covarince matrices are not required. The resulting architecture is illustrated in \figref{fig:LQGNet_scheme}.}  
%
% In practice, the complex dynamics of the underlying system determine the state-evolution model (\ref{eq:assumed_problem_a}), while the type and quality of the observations (sensors) dictate the observation model (\ref{eq:assumed_problem_b}). In some scenarios, one is likely to have access to an approximated or mismatched characterization of the underlying dynamics. Our primary focus is on scenarios where one has partial knowledge of the SS model that describes the underlying dynamics. Namely, we know (or have an approximation of) the state-evolution (transition) function $\F$ and the state-observation (emission) function $\H$. Unlike the assumptions in the classical \ac{kf}, we consider the noise statistics \textcolor{red}{$\mathbf{Q}$ and $\mathbf{R}$} unknown. Thus, we use KalmanNet to learn the \ac{kg} from data by identifying the specific computations of the \ac{kf} based on unavailable knowledge. Therefore, LQGNet integrates KalmanNet into the classical LQG flow, where Fig. \ref{fig:LQGNet_scheme} illustrates the high-level architecture.
%
%
%%%%%%%%%%%%%%%%%%%%%%%%%%
%%% Training Algorithm %%%
%%%%%%%%%%%%%%%%%%%%%%%%%%
\vspace{-0.15cm}
\subsection{Training Algorithm}\label{ssec:training}
%\vspace{-0.1cm}
The trainable parameters of \acl{ln} are the weights of the internal \ac{rnn} used to compute the \ac{kg}, denoted $\mathbf{\Theta}$. In principle, one can use the data simulator $\Gamma$ to train the state estimator separately from the control task, i.e., {to minimize the state estimate \ac{mse} in a supervised manner~\cite{KalmanNet}, or to optimize its internal predictions in an unsupervised manner~\cite{revach2021unsupervised}.} However, {inspired by \ac{rl}~\cite{SILVER2021}}, since the \ac{lqr} policy \eqref{eq:Optimal_control} uses the possibly \emph{mismatched} design matrices, we aim to train the overall system {in an \acl{e2e} manner,} based on the quadratic control objective \eqref{eq:LQR_cost}. By doing so, we leverage data to jointly cope with mismatches in both the state estimate and the regulator. 

To formulate the training loss, we use $\J_{\Gamma}(\mathbf{\Theta})$ to denote the objective \eqref{eq:LQR_cost} evaluated when applying \acl{ln} with parameters $\mathbf{\Theta}$ to control the dynamics of the simulator $\Gamma$. The random nature of the simulator indicates that restarting it multiple times yields different trajectories. We can thus formulate the regularized \ac{lqg} loss as
\begin{align}
\label{eq:lossMeasure}
\mathcal{L}_{\Gamma}\left(\mathbf{\Theta}\right) =  \J_{\Gamma}(\mathbf{\Theta}) + \gamma \cdot \Vert \mathbf{\Theta} \Vert_2^2,
\end{align}
where $\gamma>0$ is a regularization coefficient. 
By \eqref{eq:Optimal_control}, the input $\u_t$ is differentiable with respect to the state estimate $\hat{\x}_t$, which is in turn differentiable with respect to $\mathbf{\Theta}$ \cite{KalmanNet}. Consequently, the loss in \eqref{eq:lossMeasure} is differentiable, allowing to {train} \acl{ln} \acl{e2e} as a discriminative model \cite{shlezinger2022discriminative} via gradient-based learning. The resulting training procedure is summarized in Algorithm~\ref{alg:Training}. %
%%%%%%%%%%%%%%%%%
%%% Algorithm %%%
%%%%%%%%%%%%%%%%%
%
\begin{algorithm}[t]
\caption{Training LQGNet}
\label{alg:Training} 
\SetKwInOut{Initialization}{Init}
\Initialization{Randomize $\mathbf{\Theta}$ \newline Fix learning rate $\mu>0$ and  iterations $i_{\max}$}
\SetKwInOut{Input}{Input} 
\Input{Simulator  $\Gamma$}  
%  \Output{The learned parameters $\bm{\mu}^{(o)}$}
{
\For{$i = 0, 1, \ldots, i_{\max}-1$}{%
Simulate LQGNet with parameters $\mathbf{\Theta}$ using $\Gamma$\;
    Update  $\mathbf{\Theta}\leftarrow \mathbf{\Theta} - \mu\cdot \nabla_{\mathbf{\Theta}}\mathcal{L}_{\Gamma}(\mathbf{\Theta})$\;
            }
\KwRet{$\mathbf{\Theta}$}
}
\end{algorithm}
\vspace{-0.15cm}
\begin{figure*}
\begin{center}
\begin{subfigure}[h]{0.66\columnwidth}
%\centering
%\includegraphics[width=6.5cm]{LQGnet_figures/LQG_MSE_Performance_ICASSP.pdf}
\includegraphics[width= 0.92\columnwidth]{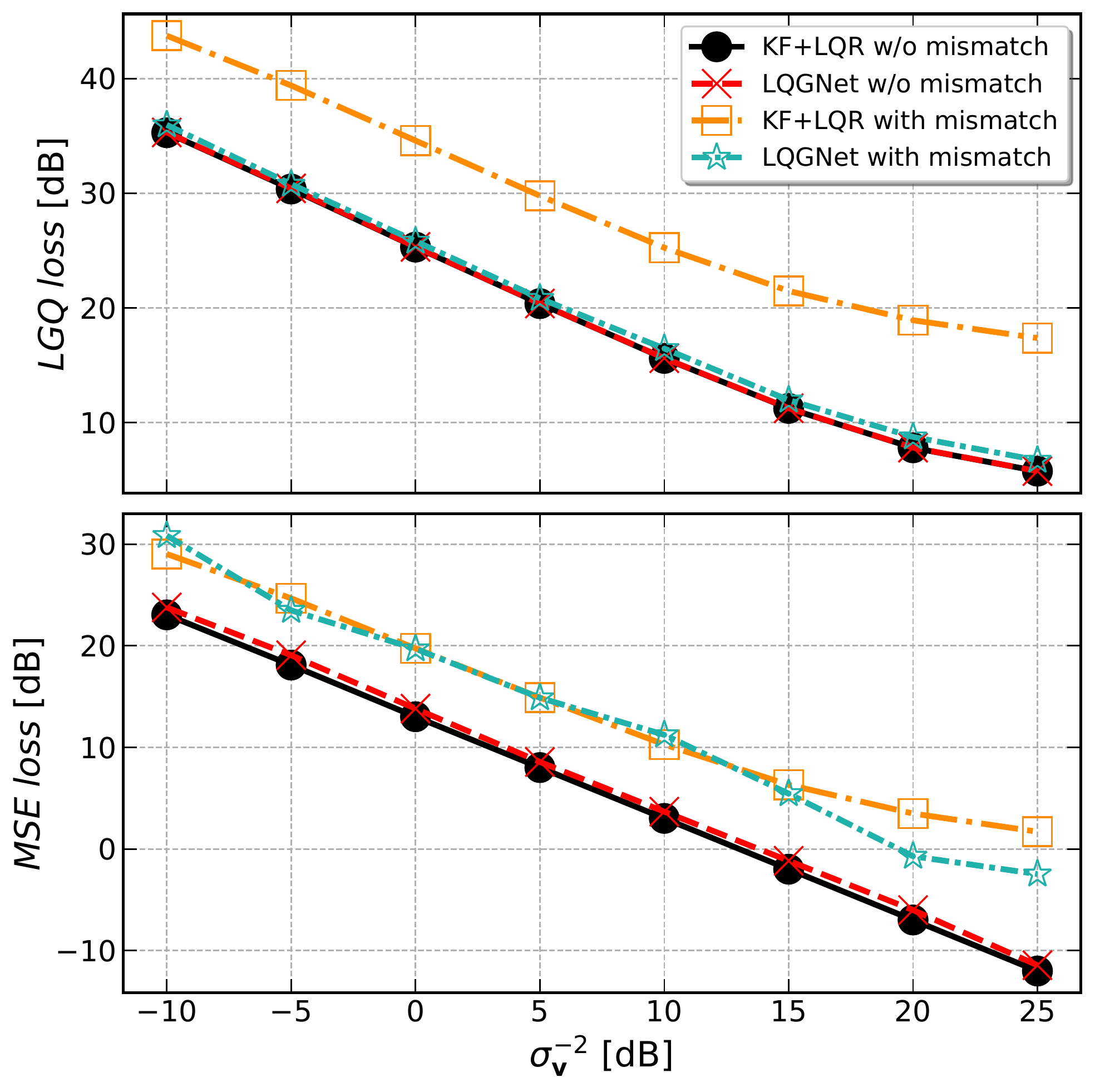}
\vspace{-1mm}
\caption{Mismatched state evolution}
\label{fig:Curve_Performance}
\end{subfigure}
\begin{subfigure}[h]{0.66\columnwidth}
%\centering
\includegraphics[width= 0.92\columnwidth]{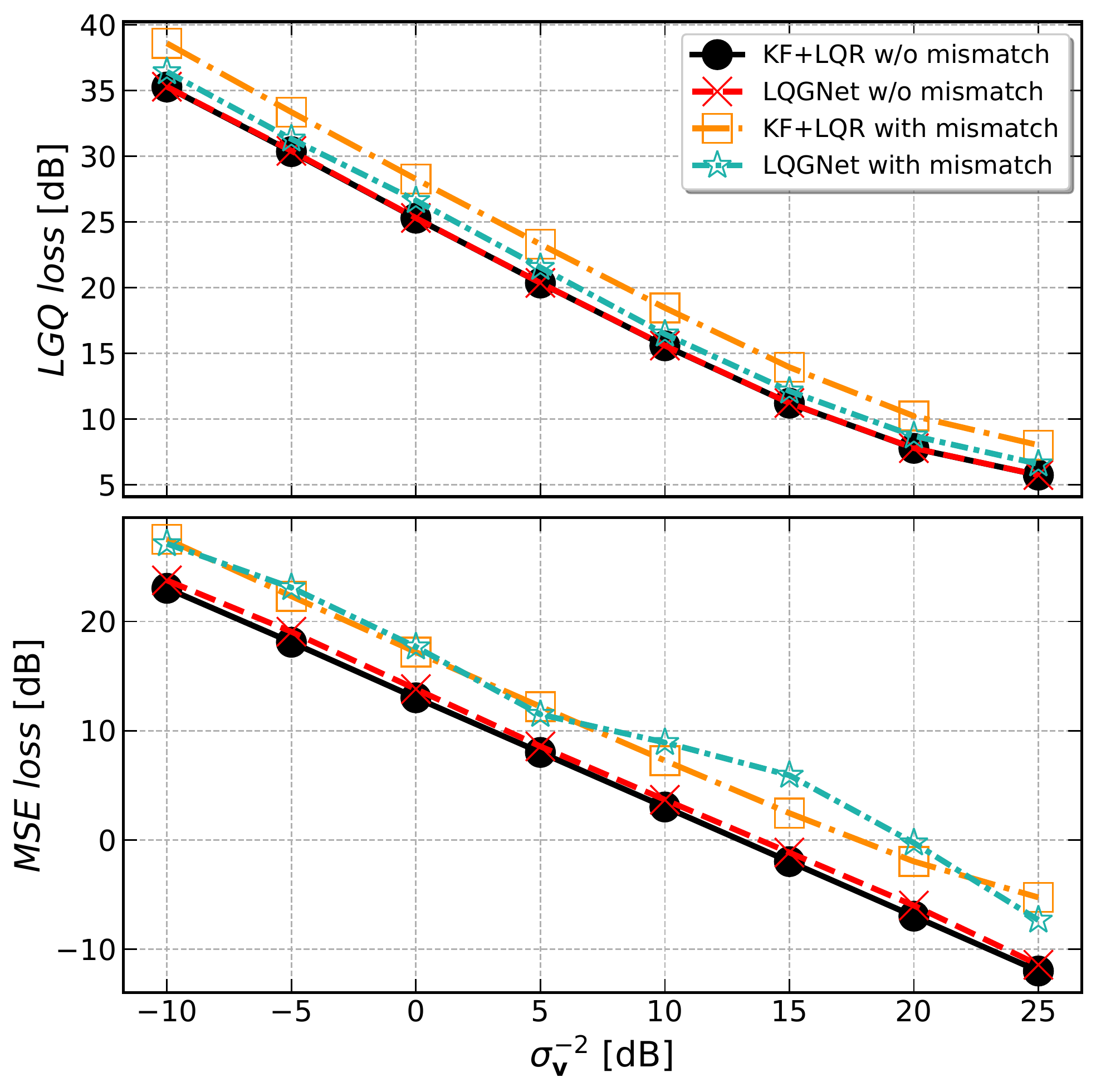}
\vspace{-1mm}
\caption{Mismatched observations}
\label{fig:MM_H}
\end{subfigure}
\begin{subfigure}[h]{0.66\columnwidth}
%\centering
\includegraphics[width= 0.92\columnwidth]{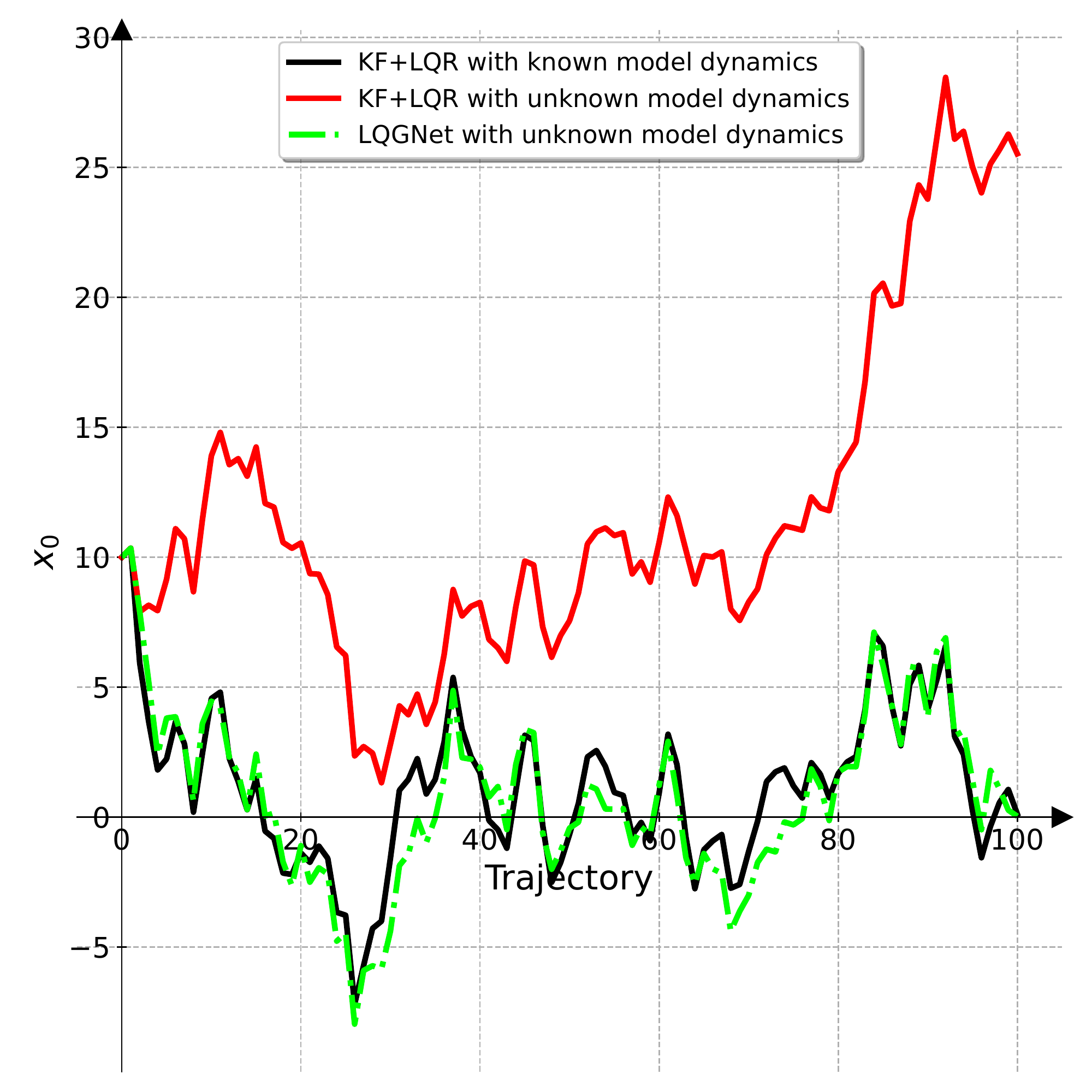}
\vspace{-0.2cm}
\caption{Trajectory along a single state entry}
\label{fig:SingleEntry_Trajectory}
\vspace{-0.2cm}
\end{subfigure}
\vspace{-1mm}
\caption{Control loss (upper) and state estimation \ac{mse} (lower) with and without mismatches.}
\label{fig:performance}
\end{center} 
\vspace{-0.4cm}
\end{figure*}
\vspace{-0.15cm}
\subsection{Discussion}\label{ssec:discussion}
%\vspace{-0.1cm}
\acl{ln} is a hybrid \acl{mb}/\acl{dd} implementation of the optimal \ac{lqg} controller. Comprising a trainable \ac{kf} and the \ac{lqr} policy, it preserves the interpretable and low-complexity operation of the \acl{mb} controller. By augmenting the \ac{kg} computation, which encapsulates the noise statistics, with an \ac{rnn}, \acl{ln} operates without imposing a model on the noise signals. The training procedure adapts \acl{ln} based on the overall control objective, facilitating coping with mismatches in the \ac{ss} model parameters. In particular, \acl{ln} can overcome uncertainty by learning an alternative state estimate that yields high performance control, as demonstrated in Section~\ref{sec:Results}. 

\acl{ln} gives rise to multiple possible extensions. For instance, while we focus on linear \ac{ss} models, the proposed design can potentially be enhanced to account for non-linear dynamics~\cite{todorov2005generalized}. Furthermore, it can be enhanced to account for alternative control objectives such as model predictive control, and utilize corresponding regulators such as emerging convex optimization control policies \cite{COCPs1}. We leave these extensions for future investigation. 

\vspace{-0.15cm}
\section{Empirical Evaluations}\label{sec:Results}
%
% are \ac{i.i.d.} Gaussian entries of identical variance, denoted $\sigma_{\w}^2$
{Next, we empirically study the performance of \acl{ln}\footnote{The source code and hyperparameters used are at \url{https://github.com/KalmanNet/LGQNet_ICASSP23}.}. We evaluate both the control objective (\ac{lqg} loss \eqref{eq:LQR_cost}) and the state estimation \ac{mse} for different levels of observation noise.} Since it  was derived from the optimal \ac{lqg} controller, we compare it with the \acl{mb} controller, i.e., \ac{kf}-\ac{lqr}, when operating with and without mismatches,  Here, the noise signals obey the \ac{ss} model in \eqref{eq:SS_model}, with diagonal covariance matrices with an identical variance, i.e., $\gvec{V}= \sigma_{\v}^2\gvec{I}$, $\gvec{W}= \sigma_{\w}^2\gvec{I}$, and $\sigma_{\w}^2= \sigma_{\v}^2$. In the non-mismatched case, the dynamics also obey the \ac{ss} model, with the design matrices
\begin{equation}
\F = \left[
\begin{array}{cc}
    1 & 1 \\
    0 & 1 \\
\end{array}
\right]; \quad 
\G = \left[
\begin{array}{c}
    0 \\
    1 \\
\end{array}
\right]; \quad 
\H = \left[
\begin{array}{cc}
    1 & 0 \\
    0 & 1 \\
\end{array}
\right].
\label{eqn:SSparameters}
\end{equation}
%
% In the mismatched case, all policies are provided with the \ac{ss} model parameters in \eqref{eqn:SSparameters}, while in the true dynamics, the state-evolution matrix $\F$ or the observation mapping $\H$ are replaced with  $\mathbf{R}_\alpha\F$ or $\mathbf{R}_\alpha\H$, respectively, with
% %
% \begin{align}
% \label{eq:rotation_matrix}
% \mathbf{R}_\alpha = 
% \left[
% \begin{array}{cc}
% \cos{(\alpha)} & -\sin{(\alpha)} \\
% \sin{(\alpha)} & \cos{(\alpha)} \\
% \end{array}
% \right],
% \end{align}    
% %
% is a mismatch rotation matrix, where we use   $\alpha = 20^\circ$.
In the mismatched case the design matrices stay the same, but in the \acl{gt} dynamics, from which the data was generated, either $\F$ or $\H$ are replaced by
\begin{align}
\label{eq:rotation_matrix}
\gvec{R}_\alpha\F,
\quad
\gvec{R}_\alpha\H,
\quad
\mathbf{R}_\alpha\!=\!
\left[
\begin{array}{cc}
\cos{(\alpha)} & -\sin{(\alpha)} \\
\sin{(\alpha)} & \cos{(\alpha)} \\
\end{array}
\right],
\end{align}  
namely, rotated with $\alpha = 20^\circ$.

The loss measures achieved by \acl{ln} compared with \acl{mb} \ac{kf}-\ac{lqr} are depicted in \figref{fig:performance} for both full and partial model information. The mismatch in \figref{fig:Curve_Performance} is in the state-evolution $\F$, and the mismatch in \figref{fig:MM_H} corresponds to the observation mapping $\H$. As expected, in the absence of mismatches, the optimal \ac{lqg} controller achieves the lowest \ac{mse} and \ac{lqg} values, as it follows the separation principle for optimality. For the same setting, \acl{ln} approaches the optimal \ac{mse} and \ac{lqg} performance. Even though \acl{ln} is trained to minimize the \ac{lqg} loss, it is not surprising that it achieves optimal performance in both \ac{lqg} and \ac{mse} metrics. Optimal control is approached because \ac{lqr} and \ac{kf} (i.e., \acl{lqe}) are dual problems \cite{KFdualLQR}, which also indicates that \acl{ln} learns to obey the separation principle when it is optimal.

Under the setting with partial (mismatched) model information, the \ac{lqg} loss associated with the \acl{ln} is notably better than its \acl{mb} counterpart. However, \acl{ln} does not accurately estimate the state here, achieving  \ac{mse} that is notably higher than the lower bound.  The learned \ac{kg} of \acl{ln} induced by training on the \ac{lqg} loss produces different state estimates than the \acl{mb} controllers. However, these estimates, despite being inaccurate, enable the mismatched \ac{lqr} controller to produce reliable inputs.
%$\frac{1}{\sigma_{\w}^2} = 15$ [dB]. For observation noise levels with $\frac{1}{\sigma_{\w}^2} \geq 15$ [dB], the \ac{mse} associated with LQGNet performs better. 

The result for the partial knowledge in the state-evolution in \figref{fig:Curve_Performance} is also reflected in  \figref{fig:SingleEntry_Trajectory}, where a single trajectory of the first entry of $\x_t$ is presented. Here, the initial state is set to $\x_0=\left[10, 0 \right]^\top$ and $\sigma_{\w}^2=\sigma_{\v}^2=0\dB$. We observe that despite the fact that \acl{ln} operates with a mismatched state-evolution matrix $\F$,  it controls the state to yield a trajectory that is closely similar to that achieved for the same setting using the optimal \ac{lqg} controller. These results demonstrate the ability of \acl{ln} to reliably learn from data to control partially known \ac{ss} models. 
%
%
% \begin{figure}
% \centering
% \includegraphics[width=8cm]{LQGnet_figures/LQG_MSE_Performance_ICASSP.pdf}
% \vspace{-0.2cm}
% \caption{Control loss (upper) and state estimation \ac{mse} (lower).}
% \label{fig:Curve_Performance}
% \vspace{-0.2cm}
% \end{figure}

% \begin{figure}
% \centering
% \includegraphics[width=8cm]{LQGnet_figures/LQG_MSE_Performance_MM_H_ICASSP.pdf}
% \vspace{-0.2cm}
% \caption{Rotation over $\H$.}
% \label{fig:MM_H}
% \vspace{-0.2cm}
% \end{figure}
%
%
% \begin{figure}
% \centering
% \includegraphics[width=7cm]{LQGnet_figures/Trajectory_Performance_ICASSAP.pdf}
% \vspace{-0.2cm}
% \caption{Trajectory along a single state entry.}
% \label{fig:SingleEntry_Trajectory}
% \vspace{-0.2cm}
% \end{figure}
%
%
% \begin{figure}
% \centering
% \includegraphics[width=8cm]{LQGnet_figures/LQG_MSE_G_Performance_ICASSP.pdf}
% \caption{Rotation over $\G$.}
% \label{fig:MM_G}
% \end{figure}
%
% \begin{figure}[t]
% \centering
% \includegraphics[width=8cm]{LQGnet_figures/LQGNet_performance_ICASSP.pdf}
% \caption{LQG loss vs SNR.}
% \label{fig:LQGNet}
% \end{figure}
%
%
% \begin{figure}[t]
% \centering
% \includegraphics[width=8cm]{LQGnet_figures/MSE_performance_ICASSP.pdf}
% \caption{MSE loss vs SNR. \textcolor{red}{put fig2 and fig3 as side by side figures}}
% \label{fig:MSE_loss}
% \end{figure}
%
%
% \begin{figure}[ht]
% \centering
% \includegraphics[width=8cm]{LQGnet_figures/Figure_1.pdf}
% \caption{Loss vs SNR.}
% \label{fig:LQGNet}
% \end{figure}
%
%\textcolor{red}{we have plenty of space left. Let's add another simulation figure.}
%\vspace{-0.2cm}
%
%%%%%%%%%%%%%%%%%%%
%%% Conclusions %%%
%%%%%%%%%%%%%%%%%%%
%
\vspace{-0.15cm}
\section{Conclusions}\label{sec:Conclusions}
%\vspace{-0.1cm}
In this work we presented \acl{ln}, a hybrid \acl{mb} and \acl{dd}  stochastic controller. Our design augments the optimal \ac{lqg} controller with a deep learning component, using the recently proposed \acl{kn} to overcome {unknown noise distributions.} We train \acl{ln} \acl{e2e}, such that it learns from data to overcome model mismatches. Our numerical study shows that \acl{ln} approaches optimal control with both a full and partial \ac{ss} model, and that it retains the simplicity and interpretability of its classic \acl{mb} counterpart while implicitly learning to cope with mismatched dynamics. 
\bibliographystyle{IEEEtran}
\bibliography{IEEEabrv, LQGNet_ICASSP}
\end{document}